\documentclass[journal,article,pdftex,moreauthors]{Definitions/mdpi}
\preto{\abstractkeywords}{\nolinenumbers}

\usepackage[acronym]{glossaries}

\newacronym{TNO}{TNO}{Netherlands Organisation for Applied Scientific Research}
\newacronym{HHS}{THUAS}{The Hague University of Applied Sciences}

\newacronym[\glslongpluralkey={Unmanned Aerial Systems}]        {UAS}   {UAS}   {Unmanned Aerial System}
\newacronym[\glslongpluralkey={Unmanned Aerial Vehicles}]       {UAV}   {UAV}   {Unmanned Aerial Vehicle}
\newacronym[\glslongpluralkey={Unmanned Underwater Vehicles}]   {UUV}   {UUV}   {Unmanned Underwater Vehicle}
\newacronym[\glslongpluralkey={Unmanned Surface Vehicles}]      {USV}   {USV}   {Unmanned Surface Vehicle}
\newacronym[\glslongpluralkey={Unmanned Ground Vehicles}]       {UGV}   {UGV}   {Unmanned Ground Vehicle}
\newacronym[\glslongpluralkey={Autonomous Underwater Vehicles}] {AUV}   {AUV}   {Autonomous Underwater Vehicle}

\newacronym[\glslongpluralkey={Remotely Operated Underwater Vehicles}]{ROV}{ROV}{Remotely Operated Underwater Vehicle}

\newacronym[\glslongpluralkey={global navigation satellite systems}]    {GNSS}  {GNSS}  {global navigation satellite systems}

\newacronym[\glslongpluralkey={Law Enforcement Agencies}]       {LEA}   {LEA}   {Law Enforcement Agency}
\newacronym[\glslongpluralkey={Counter-UAS systems}]            {C-UAS} {C-UAS} {Counter-UAS system}
\newacronym[\glslongpluralkey={Key Performance Indicators}]     {KPI}   {KPI}   {Key Performance Indicatior}
\newacronym[\glslongpluralkey={Areas of Interest}]              {AoI}   {AoI}   {Area of Interest}
\newacronym[\glslongpluralkey={Graphics Processing Units}]      {GPU}   {GPU}   {Graphics Processing Unit}

\newacronym{IMU}    {IMU}   {inertial measurement unit}
\newacronym{GPS}    {GPS}   {global positioning system}
\newacronym{RC}     {RC}    {remote controlled}

\newacronym{CNI}    {CNI}     {Critical National Infrastructure}
\newacronym{COTS}   {COTS}    {commercial off the shelf}
\newacronym{DTI}    {DTI}     {detection, tracking and identification}
\newacronym{EM}     {EM}      {electromagnetic}
\newacronym{LIDAR}  {LIDAR}   {light detection and ranging}
\newacronym{RADAR}  {RADAR}   {radio detection and ranging}
\newacronym{SONAR}  {SONAR}   {sound navigation and ranging}
\newacronym{FAR}    {FAR}     {False Alarm Rate}
\newacronym{MAR}    {MAR}     {Missed Alarm Rate}
\newacronym{POD}    {PoD}     {Probability of Detection}
\newacronym{RMS}    {RMS}     {Root Mean Square}
\newacronym{UI}     {UI}      {User Interface}
\newacronym{AI}     {AI}      {Artificial Intelligence}
\newacronym{VTOL}   {VTOL}    {vertical take-off and landing}
\newacronym{RF}     {RF}      {radio frequency}
\newacronym{IF}     {IF}      {infra-red}
\newacronym{FPV}    {FPV}     {first-person view}
\newacronym{LABEL}  {ABBREV}    {Text}

\makeglossaries

\usepackage{multirow}
\firstpage{1}
\pubvolume{1}
\issuenum{1}
\articlenumber{0}
\pubyear{2024}
\copyrightyear{2024}
\datereceived{ }
\daterevised{ } 
\dateaccepted{ }
\datepublished{ }
\hreflink{https://doi.org/} 

\newcommand{\tabletop}[0]{\hline}
\newcommand{\tablemid}[0]{\hline}
\newcommand{\tablebot}[0]{\hline}

\newcommand{\tablefont}[1]{\textbf{#1}}
\newcommand{\tablenamewidth}{5cm}
\newcommand{\tabledescriptionwidth}{12cm}
\newcommand{\tableappendixwidth}{0.5cm}
\newcommand{\labelAppendixRef}{\S}
\newcommand{\tableseparatorline}{\tablemid}

\newcommand{\tableTRUE}{\ensuremath{\bullet}}
\newcommand{\tableFALSE}{\ensuremath{\circ}}

\newcommand{\checkthree}[3]{
    \ifthenelse{\equal{#1}{1}}{\tableTRUE}{\tableFALSE}  &
    \ifthenelse{\equal{#2}{1}}{\tableTRUE}{\tableFALSE}  &
    \ifthenelse{\equal{#3}{1}}{\tableTRUE}{\tableFALSE}}

\newcommand{\checkfour}[4]{
    \ifthenelse{\equal{#1}{1}}{\tableTRUE}{\tableFALSE}  &
    \ifthenelse{\equal{#2}{1}}{\tableTRUE}{\tableFALSE}  &
    \ifthenelse{\equal{#3}{1}}{\tableTRUE}{\tableFALSE}  &
    \ifthenelse{\equal{#4}{1}}{\tableTRUE}{\tableFALSE}}

\newcommand{\checkfive}[5]{
    \ifthenelse{\equal{#1}{1}}{\tableTRUE}{\tableFALSE}  &
    \ifthenelse{\equal{#2}{1}}{\tableTRUE}{\tableFALSE}  &
    \ifthenelse{\equal{#3}{1}}{\tableTRUE}{\tableFALSE}  &
    \ifthenelse{\equal{#4}{1}}{\tableTRUE}{\tableFALSE}  &
    \ifthenelse{\equal{#5}{1}}{\tableTRUE}{\tableFALSE}}

\Title{An Objective-Driven Test Method for Comparative Performance Evaluation of Commercial DTI Solutions for Counter UAS systems}

\TitleCitation{An Objective-Driven Test Method for Comparative Performance Evaluation of Commercial DTI Solutions for Counter UAS systems}

\Author{
Ali Mohamoud\ensuremath{^{1}},
Johan van de Pol\ensuremath{^{1}},
Hanno Hildmann\ensuremath{^{1,3}}\orcidC{},
Rob van Heijster\ensuremath{^{2}},
Beatrice Masini\ensuremath{^{1}},
Martijn van den Heuvel\ensuremath{^{1}},
Amber van Keeken\ensuremath{^{1}}}

\AuthorNames{Ali Mohamoud,
Johan van de Pol,
Hanno Hildmann,
Rob van Heijster,
Beatrice Masini,
Martijn van den Heuvel,
Amber van Keeken}

\AuthorCitation{Mohamoud, A.;
Pol, J. vd.;
Hildmann, H.;
Heijster, R. v.;
Masini, B.;
Heuvel, M. vd.;
Keeken, A. v.}

\address{%
$^{1}$ \quad Automonous Systems \& Robotics (AS\&R), \gls{TNO};\\
$^{2}$ \quad Electronic Defense (ED), \glsreset{TNO}\gls{TNO};\\
$^{3}$ \quad Smart Sensor Systems lab, \gls{HHS};\\
emails: \{ali.mohamoud;
johan.vandepol;
hanno.hildmann;
rob.vanheijster;
beatrice.masini;
martijn.vandenheuvel;\\
amber.vankeeken\}@tno.nl}
\corres{Correspondence: ali.mohamoud@tno.nl}

\abstract{\glspl{UAS} or drones become more and more commercially available and cheap. The availability of these \gls{COTS} \glspl{UAS} in recent years led to potentially more threats to the traditional perimeter defense of civilian and military facilities, critical infrastructures and public events. Moreover, \glspl{LEA} find themselves confronted with the novel task of having to police the access to the lower airspace as more and more drones become available. Equally, there has been much emphasis on developing and deploying \glspl{C-UAS} with \gls{DTI} solutions. However, the capabilities of these systems are hard to benchmark. Performance claims of these systems are currently not supported by evidence. In addition, no standard test methodologies are available for these \gls{DTI} systems and different test methodologies make comparison of these systems hard or impossible.
We report on the definition, development and verification of an objective-driven test method and corresponding comparative performance evaluation for commercial \gls{DTI} solutions for \glspl{C-UAS}. The developed methodology is based on end-user scenarios that are operationally relevant. The test methodology is based on a generic \gls{DTI} system lay-out and is detailed towards detection, tracking and identification, taking into account contextual information and end-user input. The comparative performance evaluation is developed to enable the use of the methodology in a relevant environment, thereby taking into account any potential environmental aspect that might influence \gls{DTI} system performance. A comprehensive process has been outlined to develop and implement the objective-driven test method and the comparative performance evaluation. This process deals with the design of the testing system where defining the objectives the tests are addressed unambiguously. These objectives are formulated in an objective function that deals with key performance indicators for the system under test. To ensure that tests happen under comparable conditions a controlled testing environment in terms of use cases and \gls{UAS} behaviour is set up. This controlled environment covers relevant parameters and real-world aspects needed for a test scenario. Finally, such a system has to be improved and validated iteratively, meaning that once completed the test system should be applied to trials where a variety of systems are tested and compared.  The work presented in this paper has firstly been verified in a simulation environment where a number relevant scenarios were used and the output the simulation injected into the testing system. Validation of the work in a relevant environment has been done in two operational trials. The operational trial results show that the method allows for performance evaluation at component level (i.e., detection, tracking or identification component) and at system level (combinations of these components and integrated \gls{DTI} system of system solutions).}

\begin{document}
\newpage
\setcounter{tocdepth}{3}

\glsreset{COTS}
\glsreset{DTI}
\glsreset{UAV}
\section{Introduction} 
In recent years, remote controlled toys (such as cars or boats and even planes) have matured to become semi-autonomous unmanned vehicles \cite{s24072028}. Advances in virtually all sectors connected to the design, the production and the operation of such vehicles have resulted in commercially available products becoming vastly more capable (e.g., flight time, precision, autonomy, payloads) while at the same time getting cheaper and easier to operate. What used to be considered high-tech equipment just a few years ago \cite{doi/10.2788/14527} (e.g., a hovering drone with a reasonable camera providing a stable video feed of a location) is now available for very little money, shipping included, delivery to your doorstep.

In the Ukraine, \glspl{UAS} have become a \textit{consumable} resource (as opposed to a prized possession), to be used (quite literally sometimes) like an artillery shell. In the current Ukraine conflict, the majority of drones is currently operated by \gls{FPV}, meaning they are piloted (as opposed to being autonomous). With the proliferation of such \gls{FPV}-vehicles and the growing community of experienced pilots comes the reality that also our civilian airspaces are increasingly subject to such vehicles \cite{math12081250}. It is just a question of time until unmanned assets, especially \glspl{UAS} are also used for nefarious purposes \cite{doi/10.2760/969680, s24010125}.

In response, the industry has steadily advanced the field of \gls{DTI} systems. These systems, integral to both civilian and defense applications \cite{doi/10.2760/18569}, range from advanced \gls{RADAR} setups to sophisticated software solutions for \gls{UAV} oversight. Despite their technological advancements and increasing adoption, the evaluation of \gls{DTI} systems remains a complex challenge in need of objective benchmarks, comprehensive performance metrics, and adaptable testing frameworks. With commercial interests at stake, competition may be fierce and any consultant advising a customer for one and against another system (a) better have their facts straight and (b) be prepared to defend this decision against disgruntled commercial parties.

On top of that, there is the fact that, just as drones, \gls{DTI} systems can be used in a broad range of circumstances or scenarios \cite{doi/10.2788/14527}, ranging from e.g., the needs of a hobbyist airport's traffic controller over a hospital worried about the safety of their emergency helicopter's airspace to a mayor fearing for the safety of their St. Patrick's Day parade, not to mention military and defense applications. Within all these different domains a plethora of potential use-cases and scenarios can be imagined, all with their own settings and requirements. For a fair assessment of which system is best suited for a particular job the interested buyer (the user) must be able to include these details into the performance evaluation.

This paper introduces a comprehensive testing methodology tailored to the nuanced requirements of \gls{DTI} systems. By weaving together a framework that prioritizes objective-driven test methods with the comparative analysis of system performance, our approach aims to offer a versatile and rigorous means of evaluating \gls{DTI} systems. Through a meticulous integration of \glspl{KPI}, the subsequent weighting of these \glspl{KPI}, and the use of scenario-based testing, this methodology not only facilitates a nuanced assessment of \gls{DTI} capabilities but also empowers the testers to tailor the evaluation process to their specific operational needs.

\subsection{Motivation} 
As \glspl{UAS}, or drones, become more and more commercially available and cheap, \glspl{LEA} find themselves confronted with the novel task of having to police the access to the lower airspace. The availability of cheap \gls{COTS} \glspl{UAS} in recent years \cite{doi/10.2788/14527} has indeed led to potentially more threats \cite{doi/10.2760/18569} to the traditional perimeter defense of civilian and military facilities, critical infrastructures and public events \cite{systems9040079}. Equally, there has been much emphasis on developing and deploying \glspl{C-UAS} with \gls{DTI} solutions. However, the capabilities of these systems are hard to benchmark.
\newpage

These \glspl{C-UAS} provide a wide range of capabilities including \gls{DTI} functionalities. Performance claims of these systems are currently not supported by evidence. No standard test methodologies are available and different test methodologies make comparison of these systems hard or impossible. Each operational environment may require different \gls{DTI} capabilities or configurations. This may lead to \gls{DTI} systems performing differently given distinct use case scenarios and environments. There is an urgent need for the \glspl{LEA} to have an objective driven test methodology that enables comparative performance evaluation of \gls{DTI} systems. It is of paramount importance
for the \glspl{LEA} to have a such a test method including performance evaluation at their disposal.
\subsection{Scope} 
This work deals with the definition, development and verification of an objective-driven test method and comparative performance evaluation for commercial \gls{DTI} solutions for \glspl{C-UAS}. For the development of a test method that addresses relevant end-user objectives, following key elements have been taken into account \cite{MARISA2019}:
\begin{itemize}
\item A set of standard end-user defined scenarios representing a wide set of use cases (e.g., prison \& airport security, critical infrastructure protection, border security);
\item Operational end-user needs for specific mitigation techniques to counter \glspl{UAS}; and 
\item Functional and performance requirements in close collaboration with the end-users.
\end{itemize}

Based on these elements, an integral test method including comparative performance evaluation is developed that allows for a fair comparison between different \gls{DTI} systems. This test method has been verified during three user-scripted demonstration trials. In doing so, the test method fosters a better understanding among \glspl{LEA} of the capabilities and performance of \gls{DTI} systems in relevant operational use cases. In addition, relevant feedback from the\glspl{LEA} can be integrated into the test methodology.

\subsection{Contributions of this paper} 
The test method we propose, building upon the operational needs and the functional requirements,  enables testing and performance evaluation of \gls{DTI} systems. Indeed the test method allows for performance evaluation at component level (i.e., detection, tracking or identification component) and at system level (combinations of these components and integrated \gls{DTI} system of system solutions). Moreover the test method including performance evaluation takes the following main elements into account:
\begin{itemize}
\item Development of a test method for testing of integrated \gls{DTI} systems and their sub systems under realistic conditions and using relevant end-user defined scenarios.
\item A method that enables the end-users to evaluate a \gls{DTI} system and thereby helps addressing whether the system meets operational needs and requirements from the end-user perspective.
\item A future-proof methodology which means that the test methodology can be adjusted to test and evaluate future \gls{DTI} systems.
\item A black-box validation approach towards \gls{DTI} systems to make the test methodology independent of the actual design of \gls{DTI} systems.
\end{itemize}
%
\subsection{Paper Structure} 
\begin{figure}[ht]
\centering
\includegraphics[width=12cm]{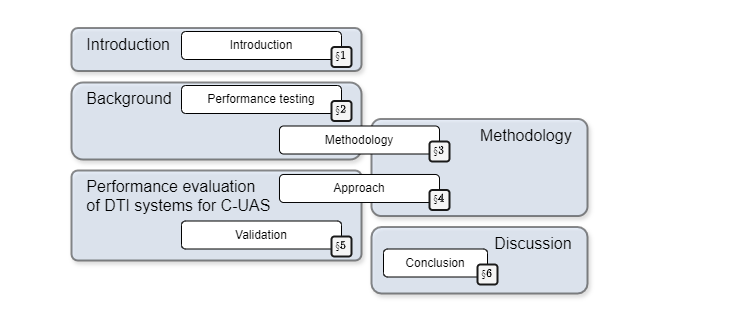}
\caption{A visual guide of the paper, outlining the relation of the sections to one another.}
\label{fig:paper.overview}
\end{figure}

Figure \ref{fig:paper.overview} provides a visual overview over the paper. We start by providing the necessary background with regard to objective driven and comparative performance testing in Section \ref{sec:PerformanceTesting}. We discuss building blocks of \gls{DTI} systems and the methodology in Section \ref{sec:MethDTI4CUAS}. This continues into Section \ref{sec:COURAGEOUS} where we report on our approach to realize objective-driven, comparative performance evaluation for commercial DTI solutions in detail. Section \ref{sec:Validation} touches on how we validated our approach and Section \ref{sec:DiscussionFutureWork} closes with a conclusion and some reflections. Formal details about the metrics used are provided in the appendices.

\section{Objective driven and comparative performance testing}\label{sec:PerformanceTesting}
\subsection{Introduction} 
When designing a performance evaluation procedure for a specific class of commercial systems, two fundamental concerns should be addressed: the process should be objective driven and the evaluation should be comparative. Both are necessary to ensure that the so evaluated systems are compared in an efficient and effective manner that is suitable to reveal whether these systems meet relevant operational needs and to make informed (e.g., purchasing) decisions towards a specific customer or end-user need.
\subsubsection{Objective driven test methods} 
Objective-driven test methods are structured around clearly defined goals and outcomes that a product, system, or process promises to achieve. This begins with the identification of the specific objectives the end-user has in mind, examples for this include
assessing the functionality of a system,
evaluating the system's performance,
verifying system reliability, or
stress-testing security aspects of a system.
Depending on the focus of the performance evaluation, this should then guide the subsequent design and development of test criteria and procedures as well as the definition of performance metrics. By focusing on the objectives of the evaluation process from the end-user perspective we can ensure that the outcomes align with the system's intended purpose and the end-user requirements.

An objective driven test method should first identify the key functional requirements of the systems to be tested (and compared). It should also consider appropriate performance benchmarks for the application. The key functional requirements are derived from relevant operational needs. In order to test different objectives relatively independently test scenarios should be designed. By subjecting a system to these scenarios all requirements and critical features can be evaluated. In addition, the decision to use objective driven performance evaluation facilitates a transparent process that can be understood and agreed upon by developers, stakeholders and end-users.
\subsubsection{Comparative performance evaluation} 
In comparative testing the performances of a number of test subjects or systems are compared, using a standardized set of criteria. It creates, in fact, a rating or an ordering of the tested entities depending on clearly defined operational needs; this allows for the results to be used directly for informed decisions with regard to the choice or purchase of a system over the others. Comparative testing can be used to facilitate a transparent, fair and unbiased comparison of the tested entities by subjecting them to identical tests and test situations and by doing so under controlled conditions.

\newpage
By designing scenarios and test cases that reflect the intended deployment of the system in the tester's specific use-case the performance comparison does compare systems with a focus on the aspects relevant to the end-user. In many cases, the ultimate decision for a system against another will be a trade-off, meaning that it is often not the case that a single system outperforms all others in all aspects and conditions. It is very likely that performance in one category comes with either a detrimental impact on another or with an excessive high cost. Often one does not need the best system at all cost but simply a system that meets the user requirements. Comparative testing enables the tester to choose from those options which yield the highest return, meaning deliver the desired service while all relevant aspects (e.g., performance, cost) are taken into account.
\subsubsection{Objective driven, comparative performance testing} 
As discussed above, both objective driven testing as well as comparative performance evaluation have significant benefits when comparing systems, software, or processes. While the former focuses on assessing whether specific considerations by the tester are met, the latter compares the performance of multiple systems against a set of tester defined benchmarks. Both methodologies contribute significantly to the making of informed decisions, however, their differences in focus and application underscore the importance of using them carefully to complement each other.

Objective driven testing is particularly relevant in early stages of a project as it forces the tester to define the objectives of a specific test at the outset. This inherently goal-oriented approach therefore drives clarity and facilitates transparent goals. It is introspective in that it focuses on the system's capability to meet its intended purpose. Comparative testing on the other hand is entirely concerned with the measurable and quantifiable performance of a system in certain conditions as it will compare different systems in the basis of these metrics. While the former ascertains that a system can indeed perform the services expected from it, the latter compares the degree with which these services are performed across systems.

This is by no means to suggest that these two approaches are mutually exclusive, far from it: both can be used in combination with great effect. For example, when designing a process to facilitate the testing of a class of systems, the design of this process should be objective driven and result in a comparative performance evaluation.
\begin{figure}[h]
\centering
\includegraphics[width=5.5cm]{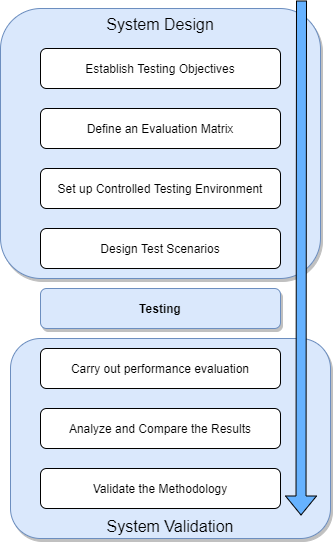}
\caption{Important process steps for comparative performance evaluation.}
\label{fig:system.design}
\end{figure}

\subsection{Important process stages} 
Designing and developing a comparative performance evaluation system happens over three high level stages: the design of the testing system, the actual use of the system (e.g., by means of an operational trial) and the validation of the system; cf. Figure \ref{fig:system.design}.

With regard to the design of such a system or process, we already discussed above the need for clearly stated objectives to unambiguously define what the goals of the testing process are. With that in place, we can discuss the metrics we want to use. These metrics must cover all aspects the tester (user) considers relevant for their eventual choice. Examples for such metrics are e.g., system performance and reliability, used hardware, response time of sensors, ease of installation, ease of use or cost (both, OPEX and CAPEX).

To ensure that tests happen under comparable conditions a controlled testing environment needs to be set up. This controlled environment should have all relevant parameters and real-world aspects needed for a test scenario (of which there are a number) covered.
The testing scenarios should be generic and abstract as well as specific enough to collectively cover the breadth of real-world scenarios.
Note that scenarios should be designed to be tunable to specific needs and interests. For example, a scenario may focus on a train station, in and contain an abstract representation of such a building. It should be tunable to a small station during peak traffic as well as to a major station in the middle of the night.

With the above in place, a number of commercially competing products can be subjected to tests and their resulting performances can be compared. Note that preferences for specific metrics over others (maybe reliability is more important than cost-effectiveness) are expressed by the end-user and are to be reflected in an objective function which is applied to all results for each tested system.

Finally, such a comparative test and evaluation methodology has to be improved and validated iteratively, meaning that once completed the system should be applied to trials where a variety of systems are tested and compared. Emphasis should be placed on the correctness of the outcome and the transparency of the process as well as the easy of use.
\subsection{System Design Principles} \label{subsec:Meth:DesignPrinciples}
Underlying the work are some fundamental principles and criteria. These are: comprehensiveness, fairness, and adaptability as well as usability and accessibility.

In order to be broadly applicable, the evaluation procedure must be \textbf{comprehensive} in that it is capable of evaluating a range of performance metrics and can so over a number of diverse scenarios. This should cover the application domain and spectrum of tested systems and represent all major use-cases (in abstract scenarios). The evaluation should be \textbf{impartial and void of inherent biases} (towards specific manufacturers, systems, paradigms or users) and should therefore be able to use any metric and run any scenario when testing a system; metrics and scenarios may not favor any individual system under investigation.

In an area as active and young as that of \glspl{UAS}, which is rapidly evolving, any testing system must be designed such that it can be amended to include changes in the field. \textbf{Adaptability} ensures that the evaluation system remains relevant as novel systems are introduced or when priorities change. This requires that metrics and scenarios can be updated or amended without it affecting the overall architecture of the system.

Finally, the evaluation system has to be \textbf{available} (both in access to the system as well as in ease of use thereof) to the broader community interested in the evaluation of a specific system. If access to the performance comparison is restricted to certain stakeholders then the main goal (that of providing a objective means to fairly assess the performance of systems to a specific use case) cannot be accomplished.
\subsection{Defining the objective} \label{subsec:Meth:ObjectiveFunction}
An objective function in the context of evaluating complex systems like \gls{DTI} systems quantitatively represents the user’s priorities and performance aspects they wish to focus on. This objective function's result is a weighted sum of normalized performance metrics, where each weight corresponds to the user's emphasis for that metric. 

Users begin by identifying \glspl{KPI} relevant to their operational needs, such as detection and track accuracy, and identification reliability. Each metric is then assigned a weight based on its importance to the user’s specific application or mission requirements. The evaluation system provides a user-friendly interface for this process, offering templates for common use cases while allowing custom configurations. The objective function is designed to be flexible, accommodating changes in user priorities and the introduction of new \glspl{KPI} as \gls{DTI} technologies evolve.

The overall aim is to design an evaluation system that can be tailored by a user to compare commercial systems for a set of specific use-cases and scenarios. The user must be able to tune the scenario to represent their use case and must further be able to change the scoring of the system to reflect their priorities and preferences. The latter is captured in the objective function of the performance evaluation, which quantitatively represents the user's priorities and focuses on those performance aspects the user prioritizes.

The objective function commonly uses normalized test scores for specific metrics, weighing them according to the user's preferences. This function should be configurable for the user, and the system should facilitate this in an easy and intuitive manner.

\subsection{Scenario modelling} \label{subsec:Meth:ScenarioModelling}
A number of abstract, but representative scenarios must be modelled. These have to reflect the diverse application domain of the system class to be tested, allowing for specific incarnations of a scenario that covers outlier cases as well as standard situations.

\section{A methodology to evaluate DTI solutions for counter-UAS systems} \label{sec:MethDTI4CUAS}
\subsection{Operational needs and scenario modelling} 
Relevant operational needs and requirements form the basis in defining functional requirements the \gls{DTI} systems should meet. The key underlying operational requirements can be summarized as the need to detect and counter any drone, of any size, exhibiting certain behaviour in different environmental conditions. Note that the countering \glspl{UAS} is not in the scope of this work.

Scenario modelling for evaluating complex systems like \gls{DTI} involves the creation of a comprehensive set of environments and conditions under which the systems' performances are assessed. These scenarios are designed to mimic a broad spectrum of real-world operational conditions, including various climates, terrains, \gls{UAS} malicious and neutral behaviour, and urban density levels. Each scenario is characterized by specific parameters such as weather conditions, the presence of obstacles, and the density of non-target entities (e.g., birds or non-malicious drones) to assess systems’ adaptability and reliability under varied conditions. The process incorporates input from industry experts, operational data, and historical incident reports to ensure relevance and comprehensiveness. This approach ensures that the evaluation system can test \gls{DTI} systems’ performance in controlled yet realistic and relevant conditions.

The proposed objective driven and comparative evaluation of the \gls{DTI} systems is built upon a set of generic scenarios. These high-level scenarios can be tailored for the \gls{LEA} specific needs. The scenarios are grouped into three main domains, specifically:
\begin{itemize}
  \item Sensitive Sites and \gls{CNI}
  \item Public Spaces Protection and Events
  \item Border Protection (Land and Maritime)
\end{itemize}

Please see Section \ref{subsec:COURAGEOUS:ScenarioModelling} for some more details on our scenarios.

The operational needs and requirements derived from these scenarios
are then translated into a set of functional and performance requirements for \gls{DTI} systems.

\subsection{Functional requirements} 
Using operational requirements and generic scenarios, a set of functional requirements are derived (these are not addressed in this paper). The operational needs translate into functional requirements for a \gls{DTI} system that can detect, track, identify (and classify) any drone, of any size, exhibiting certain behaviour in different relevant environmental conditions. Moreover, these functional requirements provide a baseline against which \gls{DTI} systems tested and evaluated. \gls{DTI} system functional breakdown, expected behaviour of the system and its components are essential for defining relevant set of performance evaluation metrics, cf. Section \ref{subsec:MethDTI4CUAS:DTI}.
\subsection{Unmanned aerial systems} 
\glspl{UAS} are technologically advanced machines that operate without a human operator physically present on the system. These systems can perform tasks autonomously, relying on pre-programmed instructions and algorithms, or be controlled remotely by a human operator from a distant location \cite{doi/10.2788/14527}. The key characteristic of unmanned systems is their ability to execute operations with varying degrees of independence from human intervention, which is made possible through a combination of sensors, data processing capabilities, and actuation mechanisms. In the context of developing objective driven and comparative evaluation of \gls{DTI} systems, \gls{UAS} also known as drones – may exhibit malicious behaviour and are as such deemed to pose a security threat in the \gls{C-UAS} domain.

\subsection{DTI building blocks} \label{subsec:MethDTI4CUAS:DTI}
The \gls{DTI} process is a means to create or increase situational awareness. In particular, awareness with regard to, in our case, (aerial) unmanned systems (drones) \cite{doi/10.2788/14527}. Therefore, the process can be said to pass through three stages: the initial realization that there is something \textit{there} (detection), the subsequent observation of that something over a period of time (tracking) which establishes the detected something as an object existing in the world, and, finally, the classification of this object (identification) according to some taxonomy of interest, e.g., \textit{friend} versus \textit{foe} or \textit{plane} versus \textit{bird}. This classification can also be of the type \textit{``is coming towards me''} or \textit{``the approaching object has malicious intent''}.

As described above and as shown in Figure \ref{fig:overview.DTI}, these three steps are sequential.

\subsubsection{Functional decomposition of DTI systems} 
\begin{figure}[ht]
\centering
\includegraphics[width=12cm]{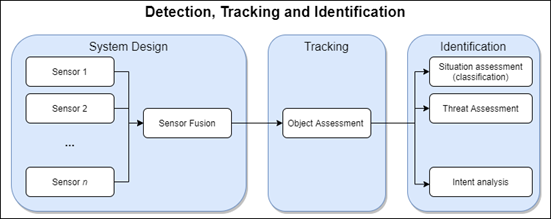}
\caption{A simplified functional overview over a \gls{DTI} system. See Sections \ref{subsec:DTI:Detection}, \ref{subsec:DTI:Tracking} and \ref{subsec:DTI:Identification} for more details on these functional components of a \gls{DTI} system, respectively.}
\label{fig:overview.DTI}
\end{figure}

The functional requirements mainly address the ability to detect a drone at a given range, translate the detections into a track over a period of time, classify the drone based on the track information and identify the drone and its potential threat \cite{AutomaticThreatEvaluation2019}. As a matter of fact, a \gls{DTI} can be decomposed into three main functional components, shown in Figure \ref{fig:overview.DTI}:
\begin{enumerate}
\item Detection functionality \cite{s24010125} (performed by detection of a signal received from the object. This can either be a signal emitted by the object, or the reflection of a signal by the object) \cite{s23177650}
\item Tracking functionality (tracking of objects – also known as object assessment – builds upon the output of the detection functionality and can be done using a single sensor or a combination of various sensors)
\item Identification functionality (includes classification of UAS and its potential threat)
\end{enumerate}

\subsubsection{Detection} \label{subsec:DTI:Detection} 
The first step in a \gls{DTI} process is detection \cite{s24072028}, where the system identifies sufficient evidence in its sensory input to infer the presence of an object, 
or at least that there is some phenomenon in the real world that has caused the sensors to measure data (as opposed to noise) \cite{s24010125}. This is a single observation and in itself not very useful. Depending on the scenario, there may be many detected objects in the sensed environment at any given time.

The technologies (sensors) required to perform detection can come from a variety of fields \cite{s24010125}: common detection sensors include
\gls{RADAR}, \gls{LIDAR}, \gls{SONAR}, \gls{IF}- or \gls{RF}-sensors, or optical cameras. The identification of evidence for the existence of an object is challenging because of the multitude of signals that can be received by a sensor in an environment. This noise has to be filtered and the remaining signal processed, the sensitivity of the sensors (as well as their range) have to be tuned to the scenario (and the object one wishes to detect) to avoid false positives while detecting as many true positives as possible (see Section \ref{subsec:MethDTI4CUAS:PerformanceMetrics:StandardMetrics}).
\subsubsection{Tracking} \label{subsec:DTI:Tracking} 
If an object has been detected once, the next step is to detect it (i.e., \textit{the same object}) again, which can be rather challenging when there are a number of objects in the environment. In a way, tracking is \textit{connecting the dots}, i.e., the recognition of a specific object so that one can label the dots representing its individual detections. Together with detections from previous moment in time these individual detections can turn into a \textit{track}. The track of a detected object represents the object's path through the environment. It might exhibit some behaviour and may be indicative of the object's nature and / or intent.

In addition, a precise track for an object possibly allows for the projection of the object's path. If, for example, the object's mass, velocity and air resistance are known then one might be able to extrapolate the resulting trajectory of the object (i.e. track prediction).

The number of detections in a track may be relevant. For example, a few dots will allow the drawing of a ragged line which would be much smoother if more dots (in-between the drawn ones) were available. If there are enough dots then the displaying only the last one with a high frequency might give the illusion of the dot moving on the screen. However, effective tracking through (quasi) continuous observations requires maintaining detection as the object moves, and to do so in a manner that enables the system to infer that the new detection and the previous one are in fact from the same object.
Consequently, tracking essentially requires detection (or sensors that facilitate detection) as well as software to process the received data and to infer which detection belongs to which object.
\subsubsection{Identification} \label{subsec:DTI:Identification} 
The identification process deals with the classification of an object based on its track and certain characteristics or attributes. This is the logical next step after detecting (becoming aware of) an object and tracking (observing) it. Classification can refer to learning more about the nature and identity of the object but may also extend to inferring the object's intent. With regard to \gls{DTI} systems, this can mean identifying
the nature (e.g., bird, drone, plane),
the type (e.g., \gls{UAV}, \gls{UGV}) or
the sub-type (e.g., quad-copter, fixed-wing) of an object, but can also extend to guessing 
the object's reason for being in the area
(e.g., hobbyist flying their drone, criminals attempting to deliver drugs to a prison or spies collecting information about the perimeter) or inferring the object's intent (such as e.g., being hostile or benign).

\begin{table}[H] 
\caption{Key aspects that the identification component of a \gls{DTI} system could address.\label{table.key.aspects.identification}}
\resizebox{\textwidth}{!}{%
\begin{tabular}{|p{5cm}|p{14cm}|}
    \tabletop
    \multicolumn{1}{|c}{\tablefont{Aspect}}	&
    \multicolumn{1}{|c|}{\tablefont{Explanation}}\\
    \tablemid
    Type Classification
    &
    Determining the type or sub-type of a detected object. Examples are distinguishing between different types of drones, birds, or other airborne objects. This might extend to recognizing characteristics that offer insight into a drone's payload (capability) or purpose (intent).\\
    \tableseparatorline
    Friend or Foe Determination
    &
    In the domain of defense, safety and security offering a high-level classification with regard to an object being friendly, hostile, or neutral (or unknown) is a crucial first step in the decision process behind planning a response.\\
    \tableseparatorline
    Behavioral Analysis
    &
    Identification can involve an analysis of the observed behaviour or a matching of observed patterns to known modi operandi. Examples are low flying drones which may want to avoid detection or devices loitering in a specific area which may indicate a surveillance activity.\\
    \tableseparatorline
    Feature Recognition
    &
    Attributing an object to some entity can be greatly improved when specific and possibly identifying features of the detected object can be observed. Examples for such features are shape, size or color but can also be signal emissions or used communication protocols. Advanced \gls{DTI} systems might use machine learning algorithms to improve feature recognition and identification accuracy over time.\\
    \tableseparatorline
    Registration and Comparison
    &
    Finally, for high value or high interest objects there may be a database with identifying features that enables a system to match an observed object uniquely to a catalogue of known objects. While maybe not the perfect example for a \gls{DTI} system, it is a known fact that navy vessels have a unique signature in their noise profile as well as their electromagnetic field. A system could e.g., uniquely identify any individual US aircraft carrier by its signatures.\\
    \tablebot
\end{tabular}
}
\end{table}

\subsection{Performance metrics} 
Performance metrics are designed to quantitatively assess a system's effectiveness, efficiency, reliability, and usability \cite{cuas2017}. Table \ref{table.metric.generic.DTI} shows example performance metrics. 
\begin{table}[H] 
\caption{Example performance metrics for \gls{DTI} systems. \label{table.metric.generic.DTI}}
\resizebox{\textwidth}{!}{%
\begin{tabular}{|p{2.8cm}|p{15.2cm}|}
    \tabletop
    \multicolumn{1}{|c}{\tablefont{Metric focus}}	&
    \multicolumn{1}{|c|}{\tablefont{Example metrics}}\\
    \tablemid
    Effectiveness
    &
    Detection range; accuracy; the system's ability to identify and track targets under various conditions.\\
    \tableseparatorline
    Efficiency
    &
    The system's resource utilization, such as energy consumption and processing time.\\
    \tableseparatorline
    Reliability
    &
    The system's operational consistency over time and under different scenarios, including its resilience to failures or environmental challenges.\\
    \tableseparatorline
    Usability
    &
    The system's ease of use, including setup time, required training, and user interface intuitiveness.\\
    \tablebot
\end{tabular}
}
\end{table}

These metrics should be standardized to allow a fair comparison across different systems but must also be customizable to reflect the specific priorities and needs of the tester. In the next section (\ref{subsec:MethDTI4CUAS:PerformanceMetrics:StandardMetrics}) we first introduce standard performance metrics before discussing the metrics of our \gls{DTI} system in specific in Sections \ref{subsec:MethDTI4CUAS:PerformanceMetrics:Detection} (Detection), \ref{subsec:MethDTI4CUAS:PerformanceMetrics:Tracking} (Tracking) and \ref{subsec:MethDTI4CUAS:PerformanceMetrics:Identification} (Identification).
\subsubsection{Standard performance metrics} \label{subsec:MethDTI4CUAS:PerformanceMetrics:StandardMetrics}
Generically speaking, and independent of the application, a few metrics are commonly used and are intuitively understandable. These can be used anywhere where there is (a) a prediction / classification, (b) a known ground truth (meaning that we do have all relevant data with regard to the actual state of affairs) and (c) more than a few results are available (meaning: where a representative number of tests have been performed). To lay the foundations for this we define true/false positives/negatives:
\begin{table}[H] 
\caption{The matrix for actual and predicted state of affairs, left generic, right an example with kittens. \label{table.predicted.versus.actual}}
\resizebox{\textwidth}{!}{%
\begin{tabular}{rcccp{1cm}rcc}
&&\multicolumn{2}{c}{\textit{Predicted}}&&&\multicolumn{2}{c}{\textit{Predicted}}\\
&&Negative&Positive&&&\textit{not a cat}&\textit{is a cat}\\
    \cline{3-4}\cline{7-8}
\multirow{2}{1cm}{\textit{Actual}}
    &Negative&\multicolumn{1}{|c}{True Negative}   &\multicolumn{1}{|c|}{False Positive}
    &&\textit{not a cat}&\multicolumn{1}{|c}{True Negative}   &\multicolumn{1}{|c|}{False Positive}   \\
    \cline{3-4}\cline{7-8}
    &Positive&\multicolumn{1}{|c}{False Negative}   &\multicolumn{1}{|c|}{True Positive}
    &&\textit{is a cat}&\multicolumn{1}{|c}{False Negative}   &\multicolumn{1}{|c|}{True Positive}   \\
    \cline{3-4}\cline{7-8}
\end{tabular}
}
\end{table}



For example, we may have 1000 pictures and know precisely which of these pictures shows a cat. We can then ask a toddler to identify all the pictures of a cat and use the responses to create 4 piles of pictures (see below and Figure \ref{table.predicted.versus.actual}). 
\begin{itemize}
\item[] \makebox[3cm][l]{\textbf{True Negative}:} images \textbf{correctly} identified as \textbf{not} having a cat in them.
\item[] \makebox[3cm][l]{\textbf{False Positive}:} images \textbf{incorrectly} identified as having a \textbf{cat} in them.
\item[] \makebox[3cm][l]{\textbf{False Negative}:} images \textbf{incorrectly} identified as \textbf{not} having a cat in them.
\item[] \makebox[3cm][l]{\textbf{True Positive}:} images \textbf{correctly} identified as having a \textbf{cat} in them.
\end{itemize}

\subsubsection{Detection} \label{subsec:MethDTI4CUAS:PerformanceMetrics:Detection}
With regard to the detection aspect of commercial \gls{DTI} systems we considered 3 metrics, namely the accuracy of the location of the detected drone (both in 2D as well as in 3D), the range ratio for detection and the precision of the detection. See Table \ref{table.metric.detection}.

\begin{table}[H] 
\caption{The 3 performance metrics chosen to assess the detection performance of a system. 
The mathematical definition for the metrics in this table are provided in Appendix \ref{Appendix:DetectionMetrics} 
(see column ``cf. \S'').\label{table.metric.detection}}
\resizebox{\textwidth}{!}{%
\begin{tabular}{|p{\tablenamewidth}|p{\tabledescriptionwidth}|p{\tableappendixwidth}|}
    \tabletop
    \multicolumn{1}{|c}{\tablefont{Metric name}}	&
    \multicolumn{1}{|c|}{\tablefont{Description}}   &
    \multicolumn{1}{|c|}{\tablefont{cf. \labelAppendixRef}}\\
    \tablemid
    Location accuracy (2D/3D)
    &
    The location accuracy of a detection representing a true object is defined as the distance between the detection and the true object. The metric is undefined for detections which do not represent a true object,
    cf. Equations
    \ref{eq:appendix:DetectionMetrics:LocationAccuracy:01},
    \ref{eq:appendix:DetectionMetrics:LocationAccuracy:02}.
    &
    \ref{app:DetectionMetrics:LocationAccuracy}\\
    \tableseparatorline
    Range ratio
    &
    The relative minimum and maximum detection range of a true object is defined as the minimum and maximum distance of the detections representing the object from the \gls{DTI} system normalized for the minimum and maximum range of the true object within the \gls{AoI} from the \gls{DTI} system,
    cf. Equations
    \ref{eq:appendix:DetectionMetrics:RangeRatio:01},
    \ref{eq:appendix:DetectionMetrics:RangeRatio:02},
    \ref{eq:appendix:DetectionMetrics:RangeRatio:03}.
    &
    \ref{app:DetectionMetrics:RangeRatio}\\
    \tableseparatorline
    Precision
    &
    The precision of detections is defined as the fraction of all detections which represent a true object,
    cf. Equation
    \ref{eq:appendix:DetectionMetrics:Precision:01}.
    &
    \ref{app:DetectionMetrics:Precision}\\
    \tableseparatorline
    Detection immediateness
    &
    The detection immediateness is defined as the difference between the time at
    which an object enters the \gls{AoI} and the time of its first detection,
    cf. Eq.
    \ref{eq:appendix:DetectionMetrics:DetectionImmediateness}.
    &
    \ref{app:DetectionMetrics:DetectionImmediateness}\\
    \tablebot
\end{tabular}
}
\end{table}



\subsubsection{Tracking} \label{subsec:MethDTI4CUAS:PerformanceMetrics:Tracking}
To evaluate the performance of a \gls{DTI} system with regard to the tracking of drones we decided upon 7 metrics: Track completeness, continuity, ambiguity and spuriousness as well as the accuracy of the track with regard to velocity and position. Finally, we included the longest track segment as a metric as well. See Table \ref{table.metric.tracking} for more details.
\begin{table}[H] 
\caption{The 7 performance metrics chosen to assess the tracking performance of a system.
The mathematical definition for the metrics this table are provided in Appendix \ref{Appendix:TrackingMetrics}.
(see column ``cf. \S'').\label{table.metric.tracking}}
\resizebox{\textwidth}{!}{%
\begin{tabular}{|p{\tablenamewidth}|p{\tabledescriptionwidth}|p{\tableappendixwidth}|}
    \tabletop
    \multicolumn{1}{|c}{\tablefont{Metric name}}	&
    \multicolumn{1}{|c|}{\tablefont{Description}}   &
    \multicolumn{1}{|c|}{\tablefont{cf. \labelAppendixRef}}\\
    \tablemid
    Track completeness
    &
    The track completeness of a true object is defined as the fraction of time in which the object is represented by at least one track,
    cf. Equation \ref{eq:appendix:TrackingMetrics:TrackCompleteness:01}.
    &
    \ref{app:TrackingMetrics:TrackCompleteness}\\
    \tableseparatorline
    Track continuity
    &
    The track continuity of a true object is defined as the total number of tracks representing the object. The metric is undefined if the true object has no tracks representing the object,
    cf. Equation \ref{eq:appendix:TrackingMetrics:TrackContinuity:01}.
    &
    \ref{app:TrackingMetrics:TrackContinuity}\\
    \tableseparatorline
    Track ambiguity
    &
    The track ambiguity of a true object is defined as the time-weighted average of the number of tracks representing the object during the time the object has at least one track representing the object. The metric is undefined if the true object has no tracks representing the object.
    Cf. Equations
    \ref{eq:appendix:TrackingMetrics:TrackAmbiguity:01},
    \ref{eq:appendix:TrackingMetrics:TrackAmbiguity:02}.
    &
    \ref{app:TrackingMetrics:TrackAmbiguity}\\
    \tableseparatorline
    Track spuriousness
    &
    The track spuriousness is defined as the time-weighted average of the number of tracks not representing a true object at that time,
    cf. Equation \ref{eq:appendix:TrackingMetrics:TrackSpuriousness:01}.
    &
    \ref{app:TrackingMetrics:TrackSpuriousness}\\
    \tableseparatorline
    Track positional accuracy
    &
    The track positional accuracy of a true object is defined as the \gls{RMS} distance between the tracks representing the object and the true object. The metric is undefined when no track represents the true object. Cf. Equations
    \ref{eq:appendix:TrackingMetrics:TrackPositionalAccuracy:01},
    \ref{eq:appendix:TrackingMetrics:TrackPositionalAccuracy:02},
    \ref{eq:appendix:TrackingMetrics:TrackPositionalAccuracy:03}.
    &
    \ref{app:TrackingMetrics:TrackPositionalAccuracy}\\
    \tableseparatorline
    Track velocity accuracy
    &
    The track velocity accuracy of a true object is defined as the \gls{RMS} velocity difference between the tracks representing the object and the true object. The metric is undefined when no track represents the true object, see Eqs.
    \ref{eq:appendix:TrackingMetrics:TrackVelocityAccuracy:01},
    \ref{eq:appendix:TrackingMetrics:TrackVelocityAccuracy:02}.
    &
    \ref{app:TrackingMetrics:TrackVelocityAccuracy}\\
    \tableseparatorline
    Longest track segment
    &
    The longest track segment of a true object is defined as the largest fraction of time in which the object was represented by the same track while being in the \gls{AoI}.
    &
    \ref{app:TrackingMetrics:LongestTrackSegment}\\
    \tableseparatorline
    Tracking immediateness
    &
    The tracking immediateness is defined as the difference between the time at which an object enters the area of interest and the time of its first associated track,
    cf. Equation
    \ref{eq:appendix:TrackingMetrics:TrackImmediateness:01}.
    &
    \ref{app:DetectionMetrics:DetectionImmediateness}\\
    \tablebot
\end{tabular}
}
\end{table}



\subsubsection{Identification} \label{subsec:MethDTI4CUAS:PerformanceMetrics:Identification}
For the identification capabilities of a \gls{DTI} system we agreed on 4 separate metrics:
F1 to capture precision and recall performance, \gls{FAR}, \gls{MAR} and \gls{POD}; these are briefly explained in Table \ref{table.metric.identification}.
\begin{table}[H] 
\glsreset{FAR}
\glsreset{MAR}
\glsreset{POD}
\caption{The 4 performance metrics chosen to assess the identification performance of a system
The mathematical definition for the metrics this table are provided in Appendix \ref{Appendix:IdentificationMetrics}
(see column ``cf. \S'').\label{table.metric.identification}}
\resizebox{\textwidth}{!}{%
\begin{tabular}{|p{\tablenamewidth}|p{\tabledescriptionwidth}|p{\tableappendixwidth}|}
    \tabletop
    \multicolumn{1}{|c}{\tablefont{Metric name}}	&
    \multicolumn{1}{|c|}{\tablefont{Description}}   &
    \multicolumn{1}{|c|}{\tablefont{cf. \labelAppendixRef}}\\
    \tablemid
    F1
    &
    The F1-score is the harmonic mean of precision and recall,
    cf. Equations
    \ref{eq:appendix:IdentificationMetrics:F1:01},
    \ref{eq:appendix:IdentificationMetrics:F1:02},
    \ref{eq:appendix:IdentificationMetrics:F1:03}.
    &
    \ref{app:IdentificationMetrics:F1}\\
    \tableseparatorline
    \glsreset{FAR}\gls{FAR}
    &
    The false alarm rate is defined as the fraction of falsely given alarms out of the total number of alarms,
    cf. Equation
    \ref{eq:appendix:IdentificationMetrics:FAR:01}.
    &
    \ref{app:IdentificationMetrics:FAR}\\
    \tableseparatorline
    \glsreset{MAR}\gls{MAR}
    &
    The missed alarm rate is defined as the fraction of alarms the \gls{DTI} did not emit out of the total number of alarms it should have emitted,
    cf. Equation
    \ref{eq:appendix:IdentificationMetrics:MAR:01}.
    &
    \ref{app:IdentificationMetrics:MAR}\\
    \tableseparatorline
    \gls{POD}
    &
    The probability of detection is defined as the number of times in which the \gls{DTI} system emits the alarm rightfully (true positives), divided by the total number of alarms it should have emitted;
    cf. Equation
    \ref{eq:appendix:IdentificationMetrics:Recall:01}.
    &
    \ref{app:IdentificationMetrics:Recall}\\
    \tableseparatorline
    Precision
    &
    The precision is defined as the fraction of all detections which represent a true object, i.e. associate with a ground truth.
    Cf. Equation
    \ref{eq:appendix:IdentificationMetrics:Precision:01}.
    &
    \ref{app:IdentificationMetrics:Precision}\\
    \tablebot
\end{tabular}
}
\end{table}



\section{An objective-driven, comparative performance evaluation for DTI solutions} \label{sec:COURAGEOUS}

\subsection{Performance evaluation pipeline} 
The performance evaluation is based on:
\begin{itemize}
\item Use of relevant end-user defined test scenarios that contain, for instance, contextual information and objects (e.g., border, \gls{CNI}) that are to be secured.
\item A set of evaluated performance metrics that, starting from the results of a test or a collection of tests, provide a score to the \gls{DTI} system under test and its components.
\item A rating that is updated every time a new test iteration has been executed.
\end{itemize}

The performance evaluation pipeline consists of two main parts, namely, Metric level (cf. Figure \ref{fig:metric.level}), and Component level (cf. Figure \ref{fig:component.level}) evaluation.
\subsubsection{The Metric Level} 
\begin{figure}[ht]
\centering
\includegraphics[width=12cm]{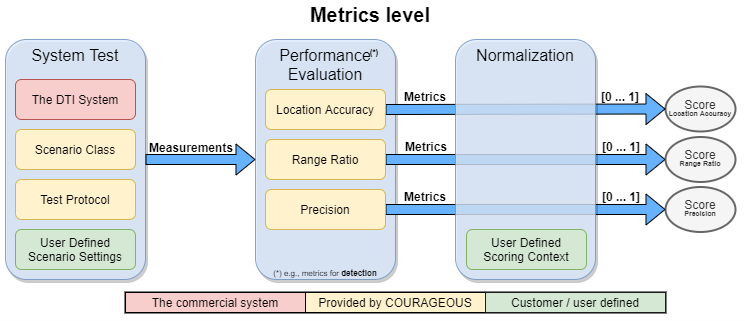}
\caption{Performance evaluation at Metric level. Note that the performance evaluation is using the three metrics for detection, cf. Table \ref{table.metric.detection}.}
\label{fig:metric.level}
\end{figure}

The metric level starts with tests being executed based on the scenarios. The measurements emanating from these tests are captured and translated into a set of metrics. At the metric level, the performance of a specific metric (e.g., detection range, track continuity) is evaluated. These metrics are then interpreted and normalized into a score. In order to know how to score the desired metric, a ‘scoring context’ can be added. This scoring context provides information about what is operationally desired for a given metric in a given context. In the end, the normalized  score between 0 and 1 is calculated, where 1 is the best possible score and 0 the worst score. See Figure \ref{fig:metric.level} for an illustration.

\subsubsection{The Component Level} 
At the component level, normalized scores calculated at the metric level are used. The component test level gathers all scores belonging to a specific functionality (e.g., detection, tracking, identification). These scores are optionally weighted based on operational end-user prioritization and then aggregated. 

\begin{figure}[ht]
\centering
\includegraphics[width=12cm]{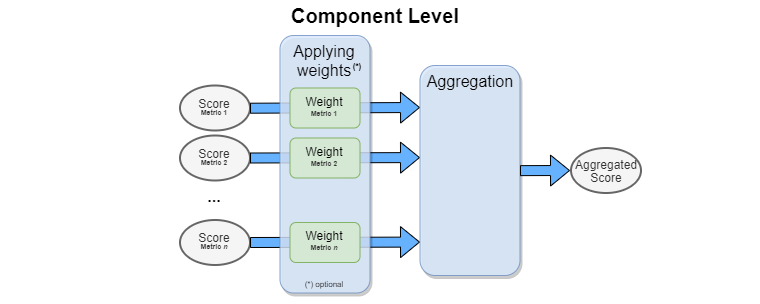}
\caption{Performance evaluation at component level.}
\label{fig:component.level}
\end{figure}

The result is an aggregated score for that specific functionality. A rating is created based on the scores of single tests, on the scores per capability and \glspl{KPI} based on operational needs. The rating is updated every time a \gls{DTI} system is tested. This is shown in Figure \ref{fig:component.level}.

The aggregated scores emanating from various component level results in the case different \gls{DTI} systems and subsystems are evaluated can be combined. In doing so, the full functionality of an integrated \gls{DTI} system, given the \gls{DTI} is comprised of a set of (sub)systems can be evaluated and compared. This combination of aggregated \gls{DTI} performance scores from this \gls{DTI} components allows for a comparative performance evaluation that enables evaluation of integrated \gls{DTI} system of systems.

\subsubsection{System Architecture} 

\begin{figure}[ht]
\centering
\includegraphics[width=13cm]{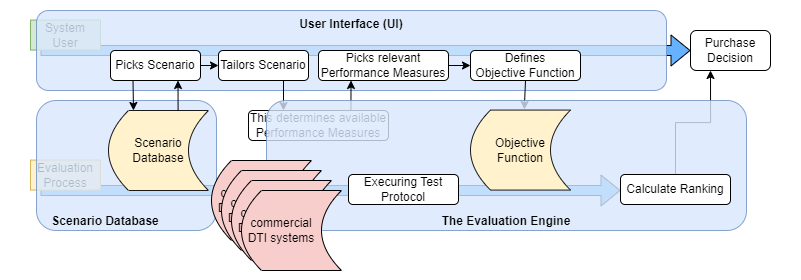}
\caption{The three main modules of the system and their interplay in the process.}
\label{fig:modules.overview}
\end{figure}

The evaluation system’s architecture is designed to be modular, scalable, and adaptable, ensuring it can accommodate a wide range of complex commercial \gls{DTI} systems and evolving evaluation needs. At its core, the architecture consists of three main components: the 
User / End-user Input, Scenario Database and the Evaluation Engine.

The User Input 
provides an accessible platform for users to define their objective functions, select or customize evaluation scenarios by configuring the metrics and corresponding weights, and view results. It is designed with usability in mind, catering to users with varying levels of technical expertise. The Evaluation Engine is the system's computational heart, processing objective functions as well as \gls{DTI} systems output based on selected scenarios and carrying out evaluation of \gls{DTI} systems though performance metrics. It utilizes a modular design, allowing new performance metrics and evaluation algorithms to be integrated as they are developed. The Scenario Database houses a comprehensive collection of pre-defined scenarios and allows for the storage of custom scenarios created by users. This component supports the system’s adaptability, enabling it to grow and evolve in response to new insights, technologies, and user requirements. 
Together, these components form a robust system capable of evaluating complex commercial \gls{DTI} systems in a fair, comprehensive, and user-driven manner.
\subsection{Defining the Objective Function}\label{subsec:COURAGEOUS:ObjectiveFunction}
The definition of an operationally relevant objective function is of vital importance for the comparative evaluation of \gls{DTI} systems. From the end-user perspective a \gls{DTI} system evaluation should be based on its ability to detect, track and identify potential malicious drones in a given \gls{AoI}. This is also of paramount importance for the subsequent mitigation measures including countering measures which are out of the scope of this work. The guiding aspects in the context of the comparative performance evaluation are testing and evaluation of integrated \gls{DTI} systems and their sub systems under realistic conditions and using relevant end-user scenarios and \glspl{KPI}. To this end, depending on the operational requirements, specific use case scenarios and the available \gls{DTI} functionality, a set of performance metrics and end-user determined weighting factors are defined. The objective function is defined such that its result is a weighted sum of normalized performance metrics where each weight corresponds to the user’s emphasis. The end-user prioritization through the use of weighting factors can be applied to a single metric, combined metrics at component level or metrics at integrated system level. The evaluation system including the objective function provides the user with the ability to formulate how well a \gls{DTI} system should cover specific operational requirements.

The objective function is designed such that it is flexible, accommodating changes in user priorities and the introduction of new \glspl{KPI} as \gls{DTI} technologies evolve.

\subsection{Scenario Modelling}\label{subsec:COURAGEOUS:ScenarioModelling}

The proposed objective driven and comparative evaluation of the \gls{DTI} systems is built upon a set of end-user defined operational needs and a set of scenarios. The scenarios are grouped into three categories:
\begin{enumerate}
\item Protection of Sensitive Sites and Critical National Infrastructure (e.g., prison, airport, nuclear plant, government building)
\item Protection of Public Spaces and Events (e.g., a stadium, outdoor events)
\item Border Protection (e.g., land border and maritime border)
\end{enumerate}

The end-user defined scenarios cover different operational needs and come with varying contextual and environmental conditions. The common denominator for these scenarios are:
\begin{itemize}
\item The definition of specific objects to be secured, including area of interest.
\item The definition of type threats (drones) that is to be expected (e.g., above ground level information, speed, size of drone and its type. Both commercial and custom-made drones are considered)
\item The expected behaviour the threat (drones) such as direction the threat is coming from, the entry into the area of interest
\item Contextual and environmental aspects (e.g., urban, suburban and rural, day and night, weather conditions and terrain)
\item The user-defined operational requirements (e.g., at which range a drone must be detected and an alarm generated. This depends on the specific scenario parameters)
\end{itemize}

\section{Validation of methodology} \label{sec:Validation}
%
%
%

A validation approach for the objective driven and comparative performance evaluation is presented here. The proposed approach is two-fold:
\begin{enumerate}
  \item A simulation-based validation approach to authenticate the developed test methodology and corresponding comparative performance evaluation in a simulation environment that facilitates sensitivity tests.
  \item A trial-based validation approach to confirm the objective driven test methodology and comparative performance evaluation in an operational environment with end-user defined relevant scenarios.
\end{enumerate}

\subsection{Simulation based validation} 
\subsubsection{Test Framework} 
\begin{figure}[ht]
\centering
\includegraphics[width=8cm]{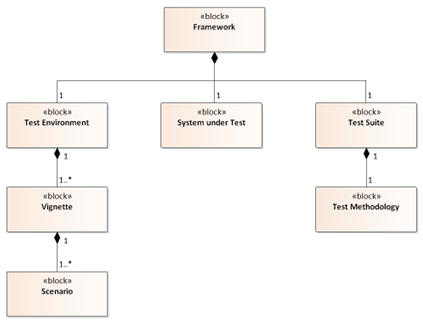}
\caption{The main components of the test framework: environment, system and test suit.}
\label{fig:test.framework}
\end{figure}

A basic concept of a test framework has been defined building upon the object driven test methodology and comparative performance evaluation presented in Section \ref{sec:COURAGEOUS}. This framework enables the validation of the proposed performance evaluation. The test framework depicted in Figure \ref{fig:test.framework} contains the following main components:
\begin{itemize}
  \item Test environment dealing with scenario's including environmental aspects (e.g., trees, birds, buildings) and objects of interest (drones).
  \item System under evaluation representing the \gls{DTI} system that is to be tested and evaluated. For the simulation-based validation, a basic
      low fidelity
      model of a \gls{DTI} system has been created with detection, tracking and identification strategies (i.e., alarm generating) capability. The purpose of the low fidelity model is to allow for verifying the comparative performance evaluation prior to its use in the trial-based validation.
  \item Test suite covering the test methodology and corresponding performance evaluation using the defined metrics.
\end{itemize}

\subsubsection{Test Environment} 
\begin{figure}[ht]
\centering
\includegraphics[width=13cm]{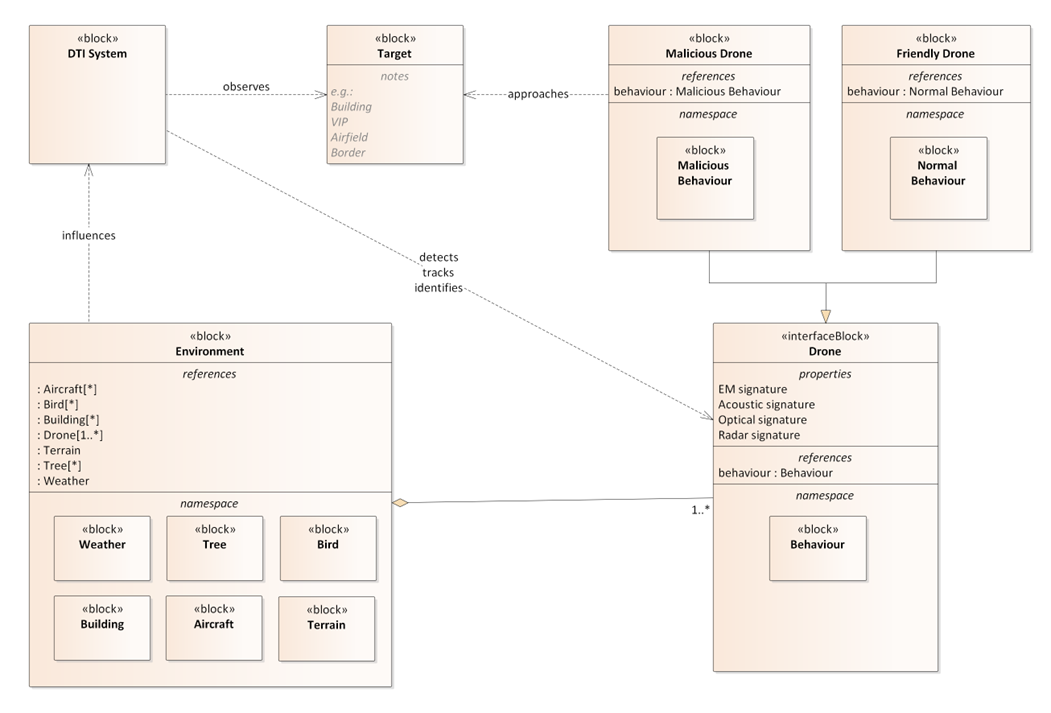}
\caption{An important characteristic of the object-driven test methodology and comparative performance evaluation is the black-box approach when it comes to \gls{DTI} systems. This means there is no need to know the internal working of the \gls{DTI} system. The \gls{DTI} black box approach means that the comparative performance evaluation does not interfere with the \gls{DTI}'s internal processing, nor does it need to measure and evaluate internal \gls{DTI} signals. The performance evaluation covers the three main functions of a \gls{DTI} system under test: Detection, Tracking and Identification.}
\label{fig:building.blocks.DTI}
\end{figure}

The test environment can be virtual (a simulation environment) or a physical test terrain where the \gls{DTI} system under test can be deployed for testing purposes (i.e., trial-based validation). An underlying requirement for the evaluation of the \gls{DTI} system under test is the inclusion of representative environment which includes objects of interest (e.g., neutral traffic and malicious drones) to be detected and classified. The concept also facilitates the storing of test data (e.g., ground truth data of objects in the environment, output of \gls{DTI} systems like detections, tracks, and identification) and logging of relevant settings (e.g., position of \gls{DTI} systems).
The test environment enables evaluation of the \gls{DTI} system in its actual environment. The interaction between the \gls{DTI} system and the main relevant aspects of the environment, shown in Figure \ref{fig:building.blocks.DTI}, contains the following components:

\begin{itemize}
  \item \gls{DTI} system representing the detection, tracking and identification system that is under test in a relevant environment.
  \item Various factors from the environment that can influence the \gls{DTI} system performance (e.g., weather, buildings, birds, aircraft, terrain and trees). Moreover, the environment can be polluted by \gls{EM} emissions (e.g., Wireless LAN, 4G).
  \item Drones can be present in the environment that exhibit certain behaviours (e.g., friendly or neutral, malicious). The drones can also generate \gls{EM} emissions. Drones are detected based on their emissions and on the signatures they have when illuminated.
  \item Location of interest depicting the area to be observed by the \gls{DTI} system which is linked to a specific scenario (e.g., border scenario, critical infrastructure).
\end{itemize}

\subsubsection{Simulation Test Suite} 
\begin{figure}[h]
\centering
\includegraphics[width=12cm]{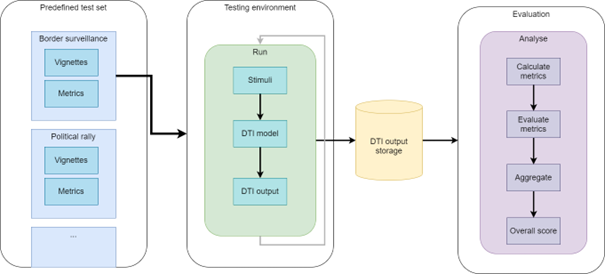}
\caption{The simulation test suite: comparative performance evaluation (right) is achieved through a rigid testing environment (middle) using predefined tests covering a collection of scenarios (left). The entire process is outlined in Figure \ref{fig:courageous.process}, the evaluation pipeline is shown in detail in Figure \ref{fig:courageous.flow}.}
\label{fig:courageous.pipeline}
\end{figure}

A test suite tailored for the simulation-based validation has been developed. This test suite given in Figure \ref{fig:courageous.pipeline} enables the execution of a number of standardized tests and generation of stimuli. After the execution of the tests, evaluation of the results of the system under test is carried out. The main components of the test suite are:
\begin{itemize}
  \item \textbf{Predefined series of tests} with a range of scenarios that are executed under controlled conditions in various vignettes. Then user-defined metrics are generated from the results of the system under test (\gls{DTI} output). These metrics are used to evaluate the system in a standardized and repeatable fashion. A test-coordinator is responsible for selecting which test is executed, preparing the test environment, starting the test itself and retrieving results.
  \item \textbf{Test environment} for generating stimuli, providing the \gls{DTI} model with necessary input, and producing \gls{DTI} output results.
  \item \textbf{Performance evaluation} process that – based on the output of the system under test during the standardized tests – comes up with a scoring of the system of interest on all tested levels.
\end{itemize}

From the predefined test sets, the scenarios to run and the stimuli to be created are known. Based on this information, multiple runs of the same test can be carried out in the test suite. From each of these runs, output of the \gls{DTI} model is generated and stored for the evaluation process. In this process, first the metrics are calculated and subsequently evaluated taking into account end-users' operational context. The evaluated metrics can be aggregated to come to an overall score for the \gls{DTI} in that specific test scenario.

\newpage

\begin{figure}[ht]
\centering
\includegraphics[width=2.2cm]{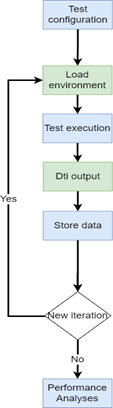}
\caption{The test pipeline.}
\label{fig:courageous.process}
\end{figure}

The steps covering the necessary elements for the validation method (cf. Fig. \ref{fig:courageous.process}) are:

\begin{itemize}
  \item Test configuration: the testing starts by loading a pre-defined test configuration.
  \item Load environment: A configuration file that defines which scenarios are run and how many iterations per scenario are required.
  \item Test execution: during an iteration, a test is executed. Execution of the test in the test environment means generating stimuli (neutral and malicious drones), obstacles (trees, buildings) and providing this information to the available \gls{DTI} models.
  \item \gls{DTI} output: Upon generation of stimuli to the \gls{DTI}, the \gls{DTI} models provide their output in terms of detections, tracking and generating of alerts. This output is then translated into metrics (\glspl{KPI}) such as detection range and track continuity.
  \item Store data: The output of the \gls{DTI} models is stored for analysis.
  \item Performance analysis: Carry out comparative performance evaluation based on the \gls{DTI} output. In case the iterations of a scenario are done, a new scenario is loaded, and new tests are performed, until all tests specified in the test configuration are handled.
\end{itemize}

\subsection{Trial-based validation} 
\subsubsection{Operational Trial Data Processing and Evaluation} 
The main objective of the trial-based validation is to authenticate the developed objective-driven test methodology and corresponding comparative performance evaluation using real world operational data. To make this possible, a data processing and analysis is defined for the trial-based validation methodology.  For the trial-based validation, operational trials are organized and executed using relevant scenarios (relevant environment, drones exhibiting both neutral and malicious behaviour) and \gls{DTI} systems. For the processing analysing of data, and subsequent evaluation of the \gls{DTI} systems, the following steps can be identified (reference in Figure \ref{fig:courageous.flow}, below):

\begin{itemize}
  \item Collection of \gls{DTI} output (e.g., detections, tracks, identification output) and ground truth data  from the flying objects (e.g., drones) from the trial data.
  \item Parsing of the collected data (i.e., \gls{DTI} output and ground truth).
  \item Pre-processing of the data (e.g., time window filtering, transformation of coordinate system and selection of \gls{AoI}).
  \item Association of detections and tracks to the ground truth data.
  \item Calculation / normalization of performance metrics. Note: there is flexibility to not normalize all of the metrics, to provide the end-user with more relevant information.
  \item Inclusion of end-user context (e.g., weighting of metrics).
  \item Generation of an aggregated score per \gls{DTI}.
\end{itemize}
\section{Analysis of Results and Discussion} \label{sec:DiscussionFutureWork}
\subsection{Discussion} 

We report here an objective-driven test methodology and a comparative performance evaluation for \gls{DTI}. The developed methodology is based on end-user scenarios that are operationally relevant. The test methodology is based on a generic \gls{DTI} system lay-out and is detailed towards \textit{detection}, \textit{tracking} and \textit{identification}, taking into account contextual information and end-user input.
The comparative performance evaluation is developed to enable the use of the methodology in a relevant environment, thereby taking into account any potential environmental aspect that might influence \gls{DTI} system performance.

The developed objective-driven test methodology and comparative performance evaluation has been verified using a two-fold approach:
\begin{itemize}
\item A simulation-based approach to validate the developed test methodology and corresponding performance evaluation. This is done in a simulation environment that facilitates carrying out sensitivity tests.
\item A trial-based validation approach to confirm the objective driven test methodology and comparative performance evaluation in an operational environment with end-user defined relevant scenarios.
\end{itemize}

For the simulation-based validation, several runs were executed to generate stimuli data as input for the test methodology and the corresponding comparative performance evaluation. Using the low fidelity \gls{DTI} model, detections (object at a certain location at a given time), track (multiple detections fused into one track over time), and alarms (in case an alarm was given by the \gls{DTI}) had been generated. Several metrics were generated from the output of the \gls{DTI} and the stored ground truth data that contains the actual positions of the simulated drones. The following metrics were used in the test runs:
\begin{itemize}
\item Range ratio
\item Location accuracy 2D
\item Track completeness
\item F1 score
\end{itemize}

The test runs were repeated for three different \gls{DTI} model strategies and different characteristics (e.g., field of view, track association capabilities). As the characteristics of the \gls{DTI} model strategies are known, validation of the performance evaluation is possible.

\begin{figure}[ht]
\centering
\includegraphics[width=9.5cm]{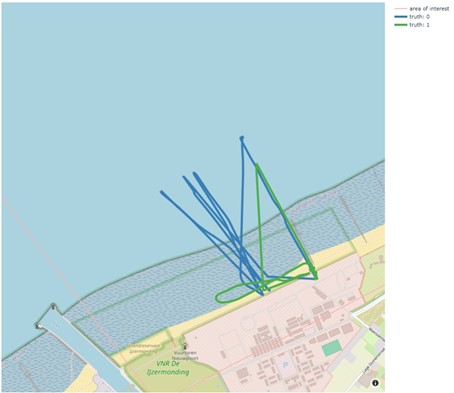}\\
\includegraphics[width=9.5cm]{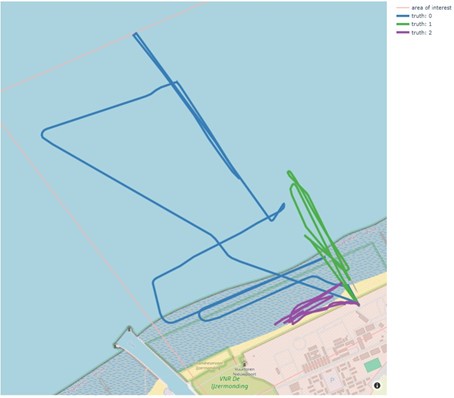}
\caption{Examples of ground truth from trials using the maritime border use case: (top) ground truth data in a scenario with two different drones in an \gls{AoI}; (bottom) ground truth data in a scenario involving three different drones.}
\label{fig:ground.truth.01}
\end{figure}

For the trial-based validation, \gls{DTI} output data (detection, tracking and identification capabilities) from two operational trials have been utilized. A number of relevant end-user defined operational scenarios have been carried out (two examples of drone flight paths – the ground truth - from trials using the maritime border use case are given in Figure \ref{fig:ground.truth.01}).

\begin{figure}[ht]
\centering
\includegraphics[width=9cm]{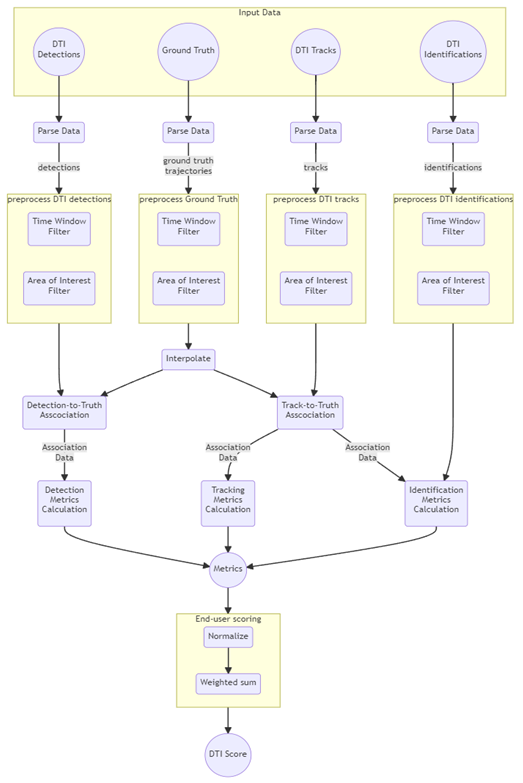}
\caption{Operational trial data processing and evaluation pipeline.}
\label{fig:courageous.flow}
\end{figure}

Note that the \gls{DTI} output obtained during the trials is, per the agreement with the participating companies, confidential and therefore not included in this report.  Various companies participated in the trials, fielding \gls{DTI} systems covering \gls{RADAR}, \textbf{EO}/\gls{IF}, \gls{RF} capabilities and combinations thereof, which were subjected to the presented test methodology in these trials. The complete processing and evaluation pipeline depicted in Figure \ref{fig:courageous.flow} has been covered. The paramount purpose of using the operational data is to validate the performance evaluation pipeline.

Using the operational data showed together with the performance evaluation pipeline, the following steps have been successfully carried out:
\begin{itemize}
\item Correct parsing of the collected data (both \gls{DTI} output such detections, tracks and identifications and drone ground truth).
\item Pre-processing of the data (e.g., time window filtering, transformation of coordinate system and selection of \gls{AoI}).
\item Association of detections and tracks to the ground truth data (\gls{DTI} track to ground-truth track association).
\item Calculation of performance metrics.
\item Inclusion of end-user context (e.g., weighting of metrics for certain metrics).
\item Generation of an score per \gls{DTI} at metric level and aggregated score at component level and system level.
\end{itemize}
More than 20 \gls{DTI} systems' output has been collected during the operational trials and processed (and example of relevant ground truth data is given in Figure \ref{fig:ground.truth.01}). Various performance metrics at different levels (metric, component, and combinations of these) have been calculated and evaluated. The trial-based validation results show the objective-driven test methodology and the comparative performance evaluation successfully process operational data, calculate performance metrics and generate (aggregated) scores for \gls{DTI} (sub)systems while taking into account the end-user input.




\printglossaries



\newpage
\authorcontributions{
Conceptualization,                      A.M.                    and R.vH.
methodology,                            A.M.            H.H.    and R.vH.
software,                                       J.vdP.                  B.M.    M.vdH.  and A.vK.
validation,                             A.M.    J.vdP.                          and M.vdH.
formal analysis,
investigation,                          A.M.    J.vdP.  H.H.    R.vH.   B.M.    M.vdH.  and A.vK.
resources,                              A.M.
data curation,                                  J.vdP.                          M.vdH.  and A.vK.
writing---original draft preparation,   A.M.            and H.H.
writing---review and editing,           A.M.    J.vdP.  H.H.    R.vH.   B.M.    M.vdH.  and A.vK.
visualization,                                  J.vdP.  H.H.                    M.vdH.  and A.vK.
supervision,                            A.M.                    R.vH.   and B.M.
project administration,                 A.M.                            and B.M.
funding acquisition,                    A.M.

All authors have read and agreed to the published version of the manuscript.}

\funding{This work has been carried out in project \textsc{COURAGEOUS}. This project has received funding from the European Union's Internal Security Fund Police (ISFP) under Grant Agreement 101034655}

\institutionalreview{“Not applicable”. No drones were harmed during the trials.}

\informedconsent{``Not applicable''.}

\dataavailability{As per the agreement with the participating companies, the performance data for specific systems under specific test conditions is considered confidential and cannot be made available publicly by the authors.}



\acknowledgments{We acknowledge the invaluable input of Kim Veltman who contributed to the COURAGEOUS project and the development of the objective-driven test methodology and comparative performance evaluation for C-UAS systems during her placement at TNO.}

\conflictsofinterest{``The authors declare no conflicts of interest''.}

\begin{adjustwidth}{-\extralength}{0cm}

\reftitle{References}


\bibliography{bibliography}

\PublishersNote{}
\end{adjustwidth}

\newpage
\appendixtitles{yes} 
\appendixstart
\appendix
\section{Detection metrics}\label{Appendix:DetectionMetrics}
\[d_{j} = \text{detection $j$}\]
\[N_{d} = \text{total number of detections}\]

\[gt_{i} = \text{ground truth $i$}\]
\[N_{\mathit{gt}} = \text{total number of ground truths}\]
{}
\[A = \text{association set between all ground truths and detections}\]
\[
A_{ij} =
	\begin{cases}
		1 & \text{if $gt_{i}$ and $d_{j}$ associate}\\
		0 & \text{otherwise}
	\end{cases}
\]
\[T(x) = \text{time stamp of the provided argument}\]

\subsection{Location accuracy (2D/3D)}\label{app:DetectionMetrics:LocationAccuracy}
The location accuracy of a true object is defined as the \gls{RMS})
distance between the locations of detections representing the object and the locations of the true object.
The location accuracy of a single ground truths are weighted using the number of associated detections, in order to put equal weight on each detection.

\begin{equation}\label{eq:appendix:DetectionMetrics:LocationAccuracy:01}
Delta_{ij} = \text{distance between $gt_{i}$ and $d_{j}$}
\end{equation}
\begin{equation}\label{eq:appendix:DetectionMetrics:LocationAccuracy:02}
\text{Location accuracy}(gt_{i}) = \sqrt{\frac
	{\sum_{j=1}^{N_{d}}A_{ij} \cdot \Delta_{ij}^{2}}
	{\sum_{j=1}^{N_{d}} A_{ij}}}
\end{equation}

\subsection{Range ratio}\label{app:DetectionMetrics:RangeRatio}
We define two ratios such that we can separately evaluate the ability of detection of a \gls{UAV} at the farthest point and at the closest point to the \GLS{DTI} system of a flight path. These metrics are called the Range Ratio Far and Range Ratio Near, respectively.


The Range Ratio Near (Far) is defined as the ratio of the nearest (farthest) detection over the range of the flight path, where the range of the flight path is the difference between the maximum and minimum distances of the ground truth to the \GLS{DTI} system. The ratios are normalized, such that the Range Ratio Near (Far) is 0 (1) if the closest (farthest) detection is equal to the nearest point of the flight path, $min_{t} (\Delta_{DTI}(\mathit{gt}_{i}, t))$ and conversely, 1 (0) if the closest (farthest) detection is equal to the farthest point of the flight path, $max_{t} (\Delta_{DTI}(\mathit{gt}_{i}, t))$.

\begin{equation}\label{eq:appendix:DetectionMetrics:RangeRatio:01}
\Delta_{DTI}(x, t) = \text{Distance between $x$ and the \GLS{DTI} sensor location at time $t$}
\end{equation}
\begin{equation}\label{eq:appendix:DetectionMetrics:RangeRatio:02}
	\text{Range ratio Near}(\mathit{gt}_{i}) = \frac
	{ max_{t} (\Delta_{DTI}(\mathit{gt}_{i}, t)) - min_{j=1}^{N_{d}} (\Delta_{DTI}(\mathit{gt}_{i}, t_{d_{j}}) \cdot A_{ij})}
	{max_{t} (\Delta_{DTI}(\mathit{gt}_{i}, t)) - min_{t} (\Delta_{DTI}(\mathit{gt}_{i}, t))}
\end{equation}
\begin{equation}\label{eq:appendix:DetectionMetrics:RangeRatio:03}
\text{Range ratio Far}(\mathit{gt}_{i}) = \frac
	{ max_{j=1}^{N_{d}} (\Delta_{DTI}(\mathit{gt}_{i}, t_{d_{j}}) \cdot A_{ij}) - min_{t} (\Delta_{DTI}(\mathit{gt}_{i}, t))}
	{max_{t} (\Delta_{DTI}(\mathit{gt}_{i}, t)) - min_{t} (\Delta_{DTI}(\mathit{gt}_{i}, t))}
\end{equation}
Note that for the distance of detection, we use the ground truth positions that are associated to the detection $d_j$ at that time $t_j$, rather than the position of the detection. This is to ensure the ratios have values between $0$ and $1$.

\subsection{Precision}\label{app:DetectionMetrics:Precision}
The precision of detections is defined as the fraction of all detections
which represent a true object.

\begin{equation}\label{eq:appendix:DetectionMetrics:Precision:01}
\text{Detection precision} = \frac{\sum_{j=1}^{N_{d}} max_{i=1}^{N_{\mathit{gt}}} Aij}{N_{d}}
\end{equation}

\subsection{Detection immediateness}\label{app:DetectionMetrics:DetectionImmediateness}
The detection immediateness is defined as the difference between the time at
which an object enters the \gls{AoI} and the time of its first detection.

\begin{equation}\label{eq:appendix:DetectionMetrics:DetectionImmediateness}
\text{Detection immediateness}(\mathit{gt}_i) = T(\mathit{gt}_i^{\mathit{start}} - min_{j=1}^{N_{d}}T(d_j)
\end{equation}

\section{Tracking metrics}\label{Appendix:TrackingMetrics}
\[t_{j} = \text{track $j$}\]
\[N_{t} = \text{total number of tracks}\]

\[\mathit{gt}_{i} = \text{ground truth $i$}\]
\[N_{\mathit{gt}} = \text{total number of ground truths}\]

\[A = \text{association between all ground truths and tracks}\]
\[
A_{ij} =
	\begin{cases}
		1 & \text{if $gt_{i}$ and $t_{j}$ associate}\\
		0 & \text{otherwise}
	\end{cases}
\]

\[R(x) = \text{time range of the provided argument}\]
\[D(x) = \text{duration of the provided argument}\]

\subsection{Track completeness}\label{app:TrackingMetrics:TrackCompleteness}
The track completeness is the duration of the true object being tracked
(by any number of unique tracks) divided by the duration of the true object
being in the area of interest.

\begin{equation}\label{eq:appendix:TrackingMetrics:TrackCompleteness:01}
\text{Track completeness}(gt_{i}) = \frac{D(\cup_{j=1}^{N_{t}} R(A_{ij}\cdot t_j))}{D(\mathit{gt}_{i})}
\end{equation}

\subsection{Track continuity}\label{app:TrackingMetrics:TrackContinuity}
The air picture is continuous when the track number assigned to an object does not change. As the air picture is represented by a collection of tracks, the track continuity is quantified with the rate of track number changes, in changes per time unit (hour). Therefore the perfect value is $0$ and higher values are worse.

We define a minimal subset of associations $A^{min} \subseteq A$ such that the coverage of the ground truth is maximized by these associated tracks.

With this, the track continuity of a true object $i$ is defined as:

\begin{equation}\label{eq:appendix:TrackingMetrics:TrackContinuity:01}
\text{Track continuity (rate)} (gt_{i})= \frac{ \sum_{j=1}^{N_t}( A^{min}_{ij}) - 1}{D(\cup_{j=1}^{N_t} R(A_{ij} \cdot t_j))}.
\end{equation}

\begin{figure}[h]
\centering
\includegraphics[width=0.6\textwidth]{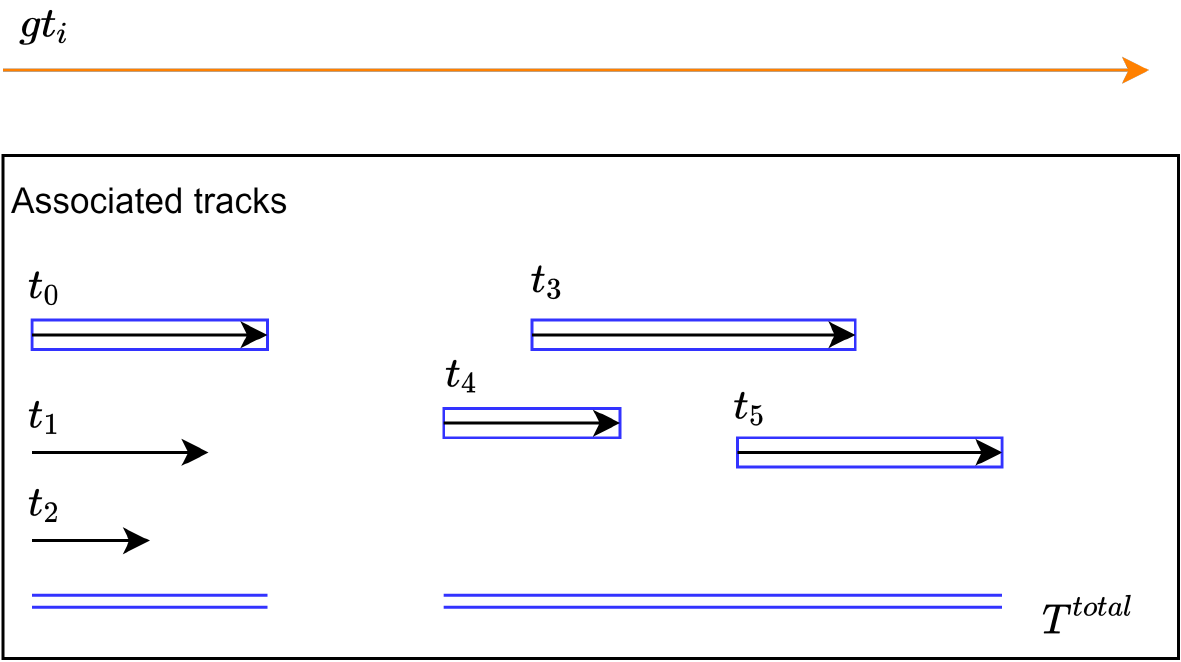}
\caption{Example of a ground truth $i$ that has 6 associated tracks. The selected tracks $t_0$, $t_3$, $t_4$ and $t_5$ (indicated by the boxes) are the minimal number of tracks ($N^{min}_i = 4$) that have the largest duration, $T^{total}_i$ (indicated by the double line). }
\label{fig:continuity-example}
\end{figure}

Figure \ref{fig:continuity-example} shows an example where 6 tracks are associated with a ground truth. The tracks that cover the most time, are $t_0$, $t_3$, $t_4$ and $t_5$. Tracks $t_1$ and $t_2$ overlap with $t_0$, thus do not increase the tracked time and therefore are not included in $T_{min}$. The track continuity is $(4-1)/\text{duration}$. If the combined duration of the subset $T_{min}$ is 1 hour the track continuity metric value is 3. We interpret this as `each hour the object was tracked, there were 3 track ID changes'.

\subsection{Track ambiguity}\label{app:TrackingMetrics:TrackAmbiguity}
The track ambiguity of a true object is defined as the time-weighted
average of the number of tracks representing the object during the time
the object has at least one track representing it. The metric is
undefined if the true object has no tracks representing it.

\begin{equation}\label{eq:appendix:TrackingMetrics:TrackAmbiguity:01}
\text{Track ambiguity}(t_{k}) = \frac{\sum_{j=1}^{N_{t}} A_{ij}(t_{k})}{N_{\mathit{gt}}(t_{k})}
\end{equation}

\begin{equation}\label{eq:appendix:TrackingMetrics:TrackAmbiguity:02}
\text{Track ambiguity} = \frac{\sum_{k=1}^{K} \text{Track ambiguity}(t_{k}) \Delta{t_{k}}}{\sum_{k=1}^{K} \Delta{t_{k}}}
\end{equation}
Note that $\Delta t_{k}$ is taken as the interval between changes in either tracks or associations, which can be different every time. The interval $[k=1, k=K]$ is determined as the time period of interest for the scenario.
When plotting, $t_{k}$ should be taken as the middle of this time period.

\subsection{Track Spuriousness}\label{app:TrackingMetrics:TrackSpuriousness}
The track spuriousness is defined as the time-weighted average of the
number of tracks not representing a true object at that time.

\begin{equation}\label{eq:appendix:TrackingMetrics:TrackSpuriousness:01}
S(t_{k}) = \frac{N_{T}(t_{k}) - N_{A}(t_{k})}{N_{T}(t_{k})}
S = \frac{\sum_{k=1}^{K}S(t_{k}) \cdot \Delta t_{k} }{\sum_{k=1}^{K}{\Delta t_{k}}}
\end{equation}
Where $S(t_{k})$ is the track spuriousness at a given time $t_{k}$, as a fraction of 1,
 $S(t_{k}) = 1$ being the worst spuriousness. $N_{T}$ and $N_{A}$ are the number of tracks
and the number of associated tracks respectively. S is the time-weighted average spuriousness S.
Note that $\Delta t_{k}$ is taken as the interval between changes in either tracks or associations, which can vary.
The interval $[k=1, k=K]$ is determined as the time period of interest for the scenario.
When plotting, $t_{k}$ should be taken as the middle of this time period.

\subsection{Track Positional Accuracy}\label{app:TrackingMetrics:TrackPositionalAccuracy}
The track positional accuracy of a true object is defined as the \gls{RMS}
distance between the tracks representing the object and the true object.
Positional errors are weighted using the average sampling rate of an
association to compensate for varying update rates between tracks.
The metric is undefined when no track represents the true object.

\begin{equation}\label{eq:appendix:TrackingMetrics:TrackPositionalAccuracy:01}
\Delta_{ij}(k) = \text{distance between $gt_{i}$ and $t_{j}$ at time $k$}
\end{equation}

\begin{equation}\label{eq:appendix:TrackingMetrics:TrackPositionalAccuracy:02}
\mathit{acc}_{ij} = \sqrt{\frac{\sum_{k=1}^{N_{k,j}} A_{ij}(k) \cdot \Delta_{ij}(k)^{2}} {N_{k,j}}}
\end{equation}

\begin{equation}\label{eq:appendix:TrackingMetrics:TrackPositionalAccuracy:03}
\text{Track positional accuracy}(gt_{i}) = \sqrt{
	\frac{\sum_{j=1}^{N_{t}} D(A_{ij}) \cdot \mathit{acc}_{ij}^{2} }
	{\sum_{j=1}^{N_{t}} D(A_{ij})}
	}
\end{equation}

\subsection{Track velocity accuracy}\label{app:TrackingMetrics:TrackVelocityAccuracy}
The track velocity accuracy of a true object is defined as the \gls{RMS}
velocity difference between the tracks representing the object and the
true object. The metric is undefined when no track represents the true
object.

\begin{equation}\label{eq:appendix:TrackingMetrics:TrackVelocityAccuracy:01}
\Delta_{ij}(k) = \text{velocity difference between $gt_{i}$ and $t_{j}$ at time $k$}
\end{equation}

\begin{equation}\label{eq:appendix:TrackingMetrics:TrackVelocityAccuracy:02}
\text{Track velocity accuracy}(\mathit{gt}_{i}) = \sqrt{
	\frac{\sum_{j=1}^{N_{t}} \sum_{k=1}^{N_{k,j}} A_{ij}(k) \cdot \Delta_{ij}(k)^{2}}
	{\sum_{j=1}^{N_{t}} \sum_{k=1}^{N_{k,j}} A_{ij}(k)}
}
\end{equation}

\subsection{Longest track segment}\label{app:TrackingMetrics:LongestTrackSegment}
The longest track segment of a true object is defined as the largest
fraction of time in which the object is represented by the same track
while being in the area of interest.

\[\text{Longest track segment}(\mathit{gt}_{i}) = \frac{max_{j=1}^{N_{t}} D(A_{ij})}{D(\mathit{gt}_{i})}\]

\subsection{Tracking immediateness}\label{app:TrackingMetrics:TrackImmediateness}
The tracking immediateness is defined as the difference between the time at
which an object enters the area of interest and the time of its first
associated track.


\begin{equation}\label{eq:appendix:TrackingMetrics:TrackImmediateness:01}
\text{Tracking immediateness}(\mathit{gt}_{i} = T(\mathit{gt}_i^{\mathit{start}}) - min_{j=1}^{N_{t}}T(t_j^{\mathit{start}})
\end{equation}

\section{Identification metrics}\label{Appendix:IdentificationMetrics}
In general, the act of identification assigns some identifying value to an object. We can consider this to be a specific ID, or a class such as `UAV', `bird' etc. or `friend' vs. `foe'.
All can be evaluated with the confusion matrix, shown in the table below:

\begin{center}
\begin{tabular}{|c|c|c|}
\hline
							& classification: A (P) & classification: Not A (N) \\
\hline
actually  A (P) 		&	True Positive (TP)	& False Negative(FN) \\
\hline
actually Not  A (N) 	& 	False Positive (FP)	& True Negative (TN) \\
\hline
\end{tabular}
\end{center}
We calculate these quadrants from the tracking data as follows.
We introduce $I$ to represent the identification of a track segment, which we refer to as being `positive' or `negative' as to hold for any identification/classification:
\[I = \text{identification status of track segments}\]
\[
I_{j} =
	\begin{cases}
		1 & \text{if $t_{j}$ is identified as positive}\\
		0 & \text{otherwise}
	\end{cases}
\]
The True Positives are calculated as the duration of the truth trajectories in which there was at least a single identified track associated to the truth, i.e.:
\begin{equation}
 \mathit{TP} =  D(\cup_{i=1}^{N_{gt}} \cup_{j=1}^{N_{t}} R(A_{ij}\cdot t_j \cdot I_{j})).
\end{equation}
The value for the False Positives constitutes to the duration of the tracks trajectories in which there was no associated truth, but the track was identified as positive, i.e.:
\begin{equation}
 \mathit{FP} = D( \cup_{j=1}^{N_t} R(t_j \cdot I_{j}) - \cup_{i=1}^{N_{gt}}\cup_{j=1}^{N_t} R(A_{ij}\cdot t_j \cdot I_{j}) )
\end{equation}
The value for the False Negatives constitutes to the duration of the truth trajectories in which there was not an \textit{identified} track associated to the truth.
This equals to the total duration of the truths minus the TP duration, i.e.:
\begin{equation}
	\mathit{FN} = \sum_{i=1}^{N_{\mathit{gt}}} D(\mathit{gt}_{i}) - \mathit{TP}
\end{equation}
The True Negatives cannot be calculated in practice, as we do not have the flight info of anything but the \glspl{UAV}. For example, we do not have the flight paths of the birds in the area, so we cannot know for sure if a track classified as 'bird' is a True Negative, or noise.

We then use these to calculate the following well-known metrics.

\subsection{F1}\label{app:IdentificationMetrics:F1}
The F1 score is a measure of classification performance, that symmetrically represents both precision and recall in one metric. Where precision and recall are defined as below:

\begin{equation}\label{eq:appendix:IdentificationMetrics:F1:01}
\text{Precision} = \frac{\mathit{TP}}{\mathit{TP} + \mathit{FP}}
\end{equation}

\begin{equation}\label{eq:appendix:IdentificationMetrics:F1:02}
\text{Recall} = \frac{\mathit{TP}}{\mathit{TP} + \mathit{FN}}
\end{equation}

The F1 is defined as:

\begin{equation}\label{eq:appendix:IdentificationMetrics:F1:03}
\text{F1} = 2 \frac{\text{precision} \cdot \text{recall}}{\text{precision} + \text{recall}} = \frac{2 \mathit{TP}}{2\mathit{TP} + \mathit{FP} + \mathit{FN}}
\end{equation}

\subsection{\gls{FAR}}\label{app:IdentificationMetrics:FAR}
The \gls{FAR} is the number of false alarms per the total number of alarms given. Also known as the False Positive Ratio.

\begin{equation}\label{eq:appendix:IdentificationMetrics:FAR:01}
\text{FAR} = \frac{\mathit{FP}}{ \mathit{FP} + \mathit{TN}}
\end{equation}
In some situations it is not possible to calculate this, as it requires exact knowledge of the negatives. If testing for the recognition of \glspl{UAV} against birds and other disturbances, it requires the flight paths of birds in the scenario, which is a practical challenge.

\subsection{\gls{MAR}}\label{app:IdentificationMetrics:MAR}
The \gls{MAR} indicates the probability of missing an alarm, where the `alarm' in this case is a classification A. It is defined as:

\begin{equation}\label{eq:appendix:IdentificationMetrics:MAR:01}
\text{MAR} = \frac{\mathit{FN}}{\mathit{TP} + \mathit{FN}}
\end{equation}

\subsection{Recall / Probability of Detection (PoD)}\label{app:IdentificationMetrics:Recall}
Complementing the \gls{MAR} is the probability of detection, which indicates the probability of sending an alarm rightfully. It is defined as the number of times in which the \GLS{DTI} system sends the alarm (classifies the object as A) rightfully, divided by the total number of objects with class A.

\begin{equation}\label{eq:appendix:IdentificationMetrics:Recall:01}
\text{PoD} = \frac{\mathit{TP}}{\mathit{TP} + \mathit{FN}}
\end{equation}

\subsection{Precision}\label{app:IdentificationMetrics:Precision}
The precision is defined as the fraction of all detections which represent a true object, i.e. associate with a ground truth.

\begin{equation}\label{eq:appendix:IdentificationMetrics:Precision:01}
\text{Precision} = \frac{\mathit{TP}}{\mathit{TP} + \mathit{FP}}
\end{equation}
\end{document}